\documentclass[12pt]{article}
\usepackage{amsmath}
\DeclareMathOperator{\tr}{tr}

\begin{document}
\author{V.G. Ksenzov\footnote{State Scientific Center Institute for Theoretical and Experimental Physics, Moscow, 117218, Russia} \!\enskip and A.I. Romanov\footnote{National Research Nuclear University MEPhI, Moscow,115409, Russia}}
\title{Dynamical Symmetry Breaking in Models with the Yukawa Interaction}
\maketitle
\begin{abstract}
We discuss models with a massless fermion and a self-interacting massive scalar field with the Yukawa interaction.
The chiral condensate and the fermion mass are calculated analytically.  It is shown that the models have a phase
transition as a function of the squared mass of the scalar field.
\end{abstract}

The dynamical chiral symmetry breaking (DCSB) plays an important role in attempts to describe different physical phenomena in the realm of elementary particles. That is why much attention has been attracted to the investigation of DCSB in different models (\cite{NJ}-\cite{BC} and references therein).

In the first paper on this subject~\cite{NJ}, Nambu and Jona-Lasinio (NJL) analyzed a specific field model in four dimensions. Later, DCSB was studied~\cite{GN} by Gross and Neveu (GN) in a $(1+1)$-dimensional spacetime in the limit of a large number of fermion flavors $N$.
These two models are similar but, in contrast to the NJL model, the GN model is a renormalizable and asymptotically free theory.
Due to these properties, the GN models are used for qualitative modeling of QCD at finite temperature and hadron density (see, e.g.~\cite{DEb} and references therein).
The relative simplicity of both models is a consequence of the quartic fermion interaction. In a real situation,
 the chiral condensate is obtained by numerical studies, especially via lattice simulations \cite{MF}.
 In these calculations, the Bank-Casher formula is often used~\cite{BC}.

In this paper we study a system of a self-interacting massive scalar and a massless fermion field with the Yukawa interaction.
In a $(1+1)$-dimensional spacetime and in the limit of a large mass of the scalar field, our model becomes equivalent to the GN model.

The studied model has a discrete symmetry that can be broken either through the Higgs mechanism (if the scalar sector makes this possible) or dynamically by means of DCSB.
The latter is characterized by the chiral condensate $\langle \bar{\psi} \psi \rangle$, which plays the role of an order parameter.
To obtain a general analytic expression for the chiral condensate, we employ the functional integral method, as described below. We find that the chiral condensate is zero when the scalar field develops a non-zero vacuum expectation value by means of the Higgs mechanism.
When there is no Higgs mechanism, the discrete symmetry is broken dynamically by the chiral condensate.
Therefore we conclude that this model has a phase transition as a function of the mass of the scalar field squared \cite{W}.

We consider a model with the Lagrangian density given by

\begin{equation}
\label{lagr}
L=L_b+L_f=\frac{1}{2} (\partial_{\mu}\phi)^2-U(\phi)+i\bar{\psi}^a\!\!\!\not\partial\psi^a-g\phi\bar{\psi}^a\psi^a,
\end{equation}
where $\phi(x)$ is a scalar field, $\psi^a (x)$ is a massless fermion field and the index $a$ runs from $1$ to $N$ ($N\gg 1$).
The potential of the scalar field,
$$U(\phi)=\frac{1}{2}m^2\phi^2+V(\phi^2),$$
includes the mass term and the self-interaction $V(\phi^2)$ of the scalar field.

Lagrangian (\ref{lagr}) is invariant under the discrete transformation

\begin{equation}
\label{trans}
\psi^a\to\gamma_5\psi^a,\quad \bar{\psi}^a\to -\bar{\psi}^a\gamma_5,\quad \phi \to -\phi.
\end{equation}
The symmetry of equation (\ref{trans}) can be broken either through the Higgs mechanism
\footnote{Note that by the Higgs mechanism in this paper we mean the spontaneous symmetry breaking at classical level,
 i.e.\ by the potential $U(\phi)$ without quantum corrections} or by the chiral condensate.
 In order to see how these two possibilities are realized, we need to calculate the chiral condensate in the model (\ref{lagr}).

The chiral condensate is given by the functional integral
\begin{equation}
\label{cond}
\langle g \bar{\psi}^a \psi ^a\rangle = \frac{1}{Z}\int {D \phi D \bar{\psi}^a D \psi^a g\bar{\psi}^a \psi^a e^{i \int{L(x)d^n x}}},
\end{equation}
where $Z$ is a normalization constant and $n$ is the spacetime dimension. Equation (\ref{cond}) can be rewritten as

\begin{equation}
\label{cond2}
\langle 0 |g\bar{\psi}^a\psi^a|0\rangle =\frac{1}{Z}\int {D \phi e^{i \int{L_b (x)d^n (x)}}} i \frac{\delta}{\delta \phi} \int{D \bar{\psi}^a D \psi^a e^{i \int{L_f (x)d^n (x)}}}.
\end{equation}

Integrating by parts~\cite{FadSlav}, we obtain
{\small
\begin{equation}
\label{cond3}
\langle 0 |g\bar{\psi}^a\psi^a|0\rangle =-\frac{1}{Z}\int {D \phi \left(\partial_{\mu}^2\phi+\frac{d}{d \phi}U(\phi)\right)e^{i \int{L_b (x)d^n (x)}} \int{D \bar{\psi}^a D \psi^a e^{i \int{L_f (x)d^n (x)}}}}.
\end{equation}}
The fermionic lagrangian is purely quadratic in the fields and we can therefore integrate over them, getting

\begin{equation}
\label{cond4}
\langle 0 |g\bar{\psi}^a\psi^a|0\rangle =-\frac{1}{Z}\int {D \phi \left(\partial_{\mu}^2\phi+\frac{d}{d\phi}U(\phi)\right)e^{i S}},
\end{equation}
where the effective action $S$ is

\begin{equation}
\label{action}
S = \int{L_b(x)d^nx}-iN \tr \ln(i\!\!\!\not \partial - g \phi).
\end{equation}

We calculate the chiral condensate using the method of the stationary phase.
A minimum of the effective action of the system can be reached if the effective potential and kinetic energy are minimal each on its own:

\begin{equation}
\label{condition}
U_\mathrm{eff}(\phi) = \min \mbox{ and } \partial_{\mu} \phi = 0,
\end{equation}

where $\int{U_\mathrm{eff}(\phi)d^n x}=\int{U(\phi)d^n x}+iN \tr \ln(i\!\!\!\not \partial - g \phi)$.

Let the constant scalar field $\phi_m$ satisfies the condition (\ref{condition}). We take into account the constant scalar field because the chiral condensate
is a constant therefore $\left.\frac{d}{d \phi}U(\phi)\right|_{\phi=\phi_m}$ in eq. (\ref{cond5}) must be a constant as well.
That is possible if $\phi_m$ is a constant, which must satisfies the eq. (\ref{condition}).
The factor in front of the exponent in eq. (\ref{cond4}) is fixed at the point $\phi = \phi_m$, which allows to evaluate the integral, one obtains

\begin{equation}
\label{cond5}
\langle 0 |g\bar{\psi}^a\psi^a|0\rangle =-\left.\frac{d}{d \phi}U(\phi)\right|_{\phi=\phi_m}.
\end{equation}

It should be noted that the accuracy of eq. (\ref{cond5}) is determined by that of the stationary phase approximation.
One can see from eq. (\ref{cond5}) that the chiral condensate is zero when the scalar field obtains a vacuum expectation value through the Higgs mechanism.

In order to study the properties of the chiral condensate, we need to choose a specific potential of the scalar field. We take the potential $U(\phi)$ as

\begin{equation}
\label{potential}
U(\phi) = \frac{1}{2}m^2\phi ^2+\frac{\lambda}{4!}\phi^4.
\end{equation}
In the following, we limit the consideration to the case of a $n=(1+1)$-dimensional spacetime.
In this case constants $g$ and $\lambda$ are dimensional, and it is convenient to introduce dimensionless coupling constants $$ g_0=\frac{g}{m},\quad \lambda_0=\frac{\lambda}{m^2}.$$

Rescalling the scalar field as $\bar{\phi}(x)=m\phi (x)$ (and omitting the bar in the rescaled field), we rewrite the effective action (\ref{action}) as

\begin{equation}
\label{action2}
S=\int{\left(\frac{1}{2m^2}(\partial_{\mu}\phi)^2-\frac{1}{2}\phi^2-\frac{\lambda_0}{m^2}\frac{\phi^4}{4!} \right)d^2x-iN\tr \ln (i\!\!\!\not \partial -g_0\phi)}.
\end{equation}
The stationary point $\phi_m$ is given by the solutions of
\begin{equation}
\label{eq1}
\phi_m+\frac{\lambda_0}{6m^2}\phi_m^3+N\frac{g_0^2\phi_m}{2\pi}\ln \frac{g_0^2\phi_m^2}{\Lambda^2} = 0,
\end{equation}
where $\Lambda$ is the ultraviolet cutoff.

Using the potential (\ref{potential}), the chiral condensate (\ref{cond5}) becomes
\begin{equation}
\label{condmin}
-\langle 0 |g_0\bar{\psi}^a\psi^a|0\rangle =\phi_m+\frac{\lambda_0}{6m^2}\phi_m^3.
\end{equation}
Since the chiral condensate (\ref{condmin}) is expressed in terms of odd powers $\phi_m$, the sign of the condensate is defined by the sign of $\phi_m$.

To investigate the properties of the chiral condensate we consider the following two possibilities:

i) $\lambda_0 = 0$. In this case the solutions of eq.~(\ref{eq1}) are $$g_0^2\phi_m^2 = \Lambda ^2 \exp \left( -\frac{2\pi}{g_0^2N}\right)$$
and, introducing

\begin{equation}
\label{nota}
g_0\phi_m=\mu_{-}^{\pm} = \pm \Lambda\exp\left( -\frac{\pi}{g_0^2N}\right),
\end{equation}
we get the chiral condensates (\ref{cond5}) as
\begin{equation}
\label{condmin2}
\langle 0 |g_0^2\bar{\psi}^a\psi^a|0\rangle =-\mu_{-}^{\pm}.
\end{equation}
This expression reproduces the result of Ref.~\cite{GN}.
Here we multiply on $g_0$ the chiral condensate to obtain a renormalization invariant quantity.
Indeed, the GN model is asymptotically free and the $\beta$ function in one-loop approximation is~\cite{GN}
\begin{equation}
\label{beta}
\beta(g_0)=\Lambda \frac{dg_0}{d\Lambda}=-\frac{g_0^3N}{2\pi}.
\end{equation}
Using (\ref{beta}), one can verify that $\mu_{-}^{\pm}$ does not depend on the choice of renormalization point,
so the fermionic mass $\mu_f=\mu_{-}^{\pm}=g_0\phi_m$ is a meaningful physical quantity.
If we change the sign in front of the $\phi^2$ term in eq.~(\ref{action2}), we get

\begin{equation}
\label{nota2}
g_0\phi_m = \mu_{+}^{\pm} = \pm \Lambda\exp\left( \frac{\pi}{g_0^2N}\right).
\end{equation}
In this case the scalar field is a tachyon and the fermionic mass is not a renormalization invariant quantity.
In the limit $\Lambda^2 \to \infty$, the fermionic mass becomes infinity as well. In both cases the fermionic mass is a nonperturbative quantity.

ii) $\lambda_0 \not = 0$. In this case eq.~(\ref{eq1}) becomes
\begin{equation}
\label{eq2}
\frac{\lambda_0}{6m^2}\phi_m^2+N\frac{g_0^2}{2\pi}\ln \frac{g_0^2\phi_m^2}{\mu_{-}^2} = 0.
\end{equation}
This equation does not admit analytic solutions for an arbitrary $\lambda_0$.
We will consider $\frac{\lambda_0}{g_0^2N}\ll1$, in which case one can solve eq.~(\ref{eq2}) by perturbing
around $\phi_m=\phi_0$:

\begin{equation}
\label{sol}
\phi_m=\phi_0+\frac{\lambda_0}{g_0^2N}\varphi. 
\end{equation}
Here $\phi_0$ is a solution of eq.~(\ref{eq1}) corresponding to $\lambda_0 = 0$. Substituting (\ref{sol}) in (\ref{eq2}), we get

 $$\varphi=-\frac{\pi}{6m^2}\phi_0^3,$$
\begin{equation}
\label{sol2}
\phi_m=\phi_0 \left( 1-\frac{\pi}{6m^2}\frac{\lambda_0}{g_0^2N}\phi_0^2 \right).
\end{equation}
Then from (\ref{condmin}) the chiral condensate is given by
\begin{equation}
\label{condminnew}
-\langle 0 |g_0^2\bar{\psi}^a\psi^a|0\rangle =\mu_{-}^{+}+\frac{\lambda_0}{6m^2g_0^2}\left(1-\frac{\pi}{g_0^2N} \right)(\mu_{-}^{+})^3.
\end{equation}
Here the solution $\mu_{-}^{+}$ was chosen for a definiteness.
The fermionic mass in this case yields
\begin{equation}
\label{fmass}
\mu_f=g_0\phi_m=\mu_{-}^{+} \left( 1-\frac{\pi}{6m^2}\frac{\lambda_0}{g_0^2N} (\mu_{-}^{+})^2\right).
\end{equation}
One can see that the scalar field potential destroys the renormalization invariance of the fermionic mass (\ref{fmass}).

For a tachyonic scalar field $\phi$ ($m^2 < 0$) the fermionic mass must be obtained by the Higgs mechanism, because the chiral condensate is zero.
Finally, we can see that the existence of the chiral condensate depends on the sign of $m^2$: there is a condensate for positive $m^2$ but not for negative $m^2$.
We conclude that this model has a phase transition as a function of $m^2$.
A phase transition of this kind was discussed in~\cite{W}.

To summarize, in this paper we studied dynamical chiral symmetry breaking in a model consisting of a fermion field and a self-interacting scalar field,
with the Yukawa interaction between the two fields. In our model, the discrete chiral symmetry can be broken either by the Higgs mechanism
or dynamically by a non-zero chiral condensate. It was shown that the presence of the Higgs mechanism does not allow for a non-zero
chiral condensate.
We further constrained the study to a $(1+1)$-dimensional spacetime and in the limit of large $N$.
Also, the self-interaction of the scalar field
was assumed to be treatable as perturbation. In this case analytic expressions for the chiral condensate and the fermionic mass have been obtained. 
We conclude that the model considered has a phase transition as function of $m^2$, since the chiral condensate is zero in the Higgs case ($m^2 < 0$)
and is not zero in the case $m^2 > 0$. The method proposed here makes it possible to calculate the chiral condensate for various scalar potentials
and for arbitrary spacetime dimensions.

\begin{center}
Acknowledgments
\end{center}

We are grateful to O.V. Kancheli, A.E. Kudryavtsev, Y.M. Makeenko, V.A. Lensky and M.A. Zubkov for stimulating discussions.

\end{document}